\documentclass{article}

\usepackage{arxiv}

\usepackage[utf8]{inputenc} 
\usepackage[T1]{fontenc}    
\usepackage{hyperref}       
\usepackage{url}            
\usepackage{booktabs}       
\usepackage{amsfonts}       
\usepackage{nicefrac}       
\usepackage{microtype}      
\usepackage{lipsum}		
\usepackage{graphicx}
\usepackage{doi}
\usepackage{mathptmx} 
\usepackage{helvet}   
\usepackage{courier}  
\usepackage{type1cm}  
\usepackage{makeidx}   
\usepackage{multicol}  
\usepackage[bottom]{footmisc}
\usepackage{todonotes}
\usepackage{multirow}
\usepackage{graphicx}
\usepackage{caption}
\usepackage{subcaption}
\usepackage{times}
\usepackage{longtable}
\usepackage{algorithm}
\usepackage{algpseudocode}
\usepackage[section]{placeins}

\pdfoutput=1

\algnewcommand\algorithmicforeach{\textbf{for each}}
\algdef{S}[FOR]{ForEach}[1]{\algorithmicforeach\ #1\ \algorithmicdo}

\title{Inferring Strategies from Observations in Long Iterated Prisoner's Dilemma Experiments}

\date{} 					

\author{\href{https://orcid.org/0000-0002-2380-8630}{\includegraphics[scale=0.06]{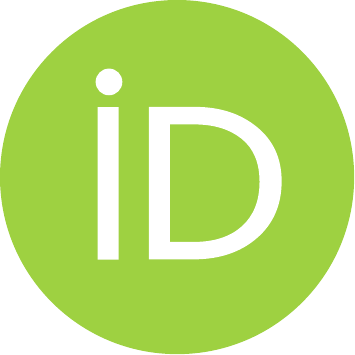}\hspace{1mm}Eladio Montero-Porras $^{1}$} \\
	\texttt{eladio.montero@vub.be} \\
	\AND
	Jelena Grujić $^{1}$\\
	\texttt{jelenagr@gmail.com} \\
	\AND
	Elias Fernandez-Domingos $^{1,2}$\\
	\texttt{elias.fernandez.domingos@vub.be} \\
	\AND
	Tom Lenaerts $^{1,2,3,4}$\\
	\texttt{tom.lenaerts@vub.be} \\
}

\hypersetup{
pdftitle={Inferring Strategies from Observations in Long Iterated Prisoner's Dilemma Experiments},
pdfsubject={cs.GT},
pdfauthor={Eladio Montero-Porras, Jelena Grujić, Elias Fernandez-Domingos, and Tom Lenaerts},
pdfkeywords={Iterated Prisoner's Dilemma, Strategies, Behavioral Experiments},
}

\begin{document}
\maketitle

$^1$: Artificial Intelligence Laboratory, Vrije Universiteit Brussel, Brussels, 1050, Belgium \\
$^2$: Machine Learning Group, Université Libre de Bruxelles, Brussels, 1050, Belgium \\
$^3$: Center for Human-Compatible AI, UC Berkeley, Berkeley, 94702, USA \\
$^4$: FARI Institute, Université Libre de Bruxelles-Vrije Universiteit Brussel, 1050 Brussels, Belgium \\ 

\begin{abstract}
While many theoretical studies have revealed the strategies that could lead to and maintain cooperation in the Iterated Prisoner's Dilemma, less is known about what human participants actually do in this game and how strategies change when being confronted with anonymous partners in each round. Previous attempts used short experiments, made different assumptions of possible strategies, and led to very different conclusions. We present here two long treatments that differ in the partner matching strategy used, i.e. fixed or shuffled partners. Here we use unsupervised methods to cluster the players based on their actions and then Hidden Markov Model to infer what are those strategies in each cluster. Analysis of the inferred strategies reveals that fixed partner interaction leads to a behavioral self-organization. Shuffled partners generate subgroups of strategies that remain entangled, apparently blocking the self-selection process that leads to fully cooperating participants in the fixed partner treatment. Analyzing the latter in more detail shows that AllC, AllD, TFT- and WSLS-like behavior can be observed. This study also reveals that long treatments are needed as experiments less than 25 rounds capture mostly the learning phase participants go through in these kinds of experiments.
\end{abstract}


\section*{Introduction}

 The emergence of cooperation in human populations has been an object of study in many different disciplines ranging from social sciences \cite{rand_human_2013, nowak_five_2006, gracia-lazaro_human_2012}, physics and complex systems \cite{santos_social_2008, perc_statistical_2017, perc_heterogeneous_2010} to biology \cite{ashlock_exploration_2006, dugatkin_guppies_1988, lombardo_mutual_1985} and computer science \cite{de_melo_social_2018, fernandez-domingos_delegation_2021}, among many others. While cooperation is widespread in human societies, its origins remain unclear, although many mechanisms have been proposed to explain its evolutionary advantage \cite{nowak_five_2006}. In fact, many social situations pose a dilemma, in which, while being cooperative towards society is beneficial for the population as a whole, free-riding on the efforts of others may generate substantial individual gains \cite{dawes_social_1980,  lange_social_2014}. While many studies have explored which individual behaviors can promote cooperation in these social dilemmas \cite{fernandez-domingos_delegation_2021, han_emergence_2013, rand_direct_2009} or aimed to find how much human decision-making aligns with those cooperative strategies \cite{rand_human_2013,baek_comparing_2016}, not so much is known about which strategies effectively can be extracted from experimental game-theoretical data, which often do not end in full cooperation. Identifying the actual decision-making schemes is essential, especially if one wants to understand why cooperation is not achieving anticipated levels, to grasp the incentives needed to promote beneficial outcomes or how artificial systems may need to be designed so that they align with human pro-social behavior.  
 
 Using a data science approach we aim to provide an answer to such questions here, focusing on a new cohort of game theoretical experiments within the well-known framework of the pairwise Iterated Prisoner's Dilemma (IPD)\cite{axelrod_evolution_1987}. In the IPD, when both players cooperate (C) in a round, they both get a reward (represented by $R$), while if one of them defects (D), and the other cooperates they get a payoff $T$ and $S$ respectively. If both defect, they both get a payoff $P$. The dilemma emerges when $T>R>P>S$, with $2R>T+S$. The payoff is accumulated by the interacting players at each round until the game ends. In this iterated version of the one-shot PD there are a plethora of possible equilibrium outcomes \cite{garcia_no_2018}, including cooperative ones \cite{fudenberg_folk_1986}.

The literature has reported on many strategies that can induce cooperation in the IPD with fixed partners. These results were produced mainly through models and simulations. In the famous first Axelrod tournament \cite{axelrod_evolution_1981}, for instance, Tit-for-Tat (TFT) emerged as the most successful strategy. TFT starts by signaling the intention to cooperate and afterwards mimics the previous action of the opponent, and it has been argued to be a good representation of reciprocal behavior in human societies. Reciprocity has been thoroughly researched as one of the most important mechanisms to favour cooperation \cite{trivers_evolution_1971, nowak_five_2006, rand_human_2013}. More generally, conditionally cooperative strategies appear to generate the right set of opportunities for cooperative behavior to spread in large populations \cite{nowak_stochastic_1990, reuben_revisiting_2012, fernandez-domingos_timing_2020}, highlighting the importance of context and past experiences in the effectiveness of cooperative strategies. Reciprocity thus justifies, from an evolutionary perspective, the existence of altruistic and pro-social behaviors \cite{gurven_collective_2008}.

However successful TFT was to explain aspects such as direct reciprocity and conditional cooperation, this strategy is known to dissolve into mutual defection in the presence of execution errors, i.e. when participants fail to keep the implicit cooperative agreement at a given round \cite{wu_how_1995}, especially in heterogeneous populations \cite{nowak_tit_1992}. Attempts to repair this flaw lead to the introduction of Generous-Tit-for-Tat (GTFT) \cite{wedekind_human_1996} which cooperates if the opponent cooperates, but if the opponent defects it sometimes “forgives” and continues to cooperate. This strategy is also a reactive strategy, but unlike TFT, it is stochastic, because the player's next action is now given with some probability. Yet, GTFT has its own problems leading to the introduction of the Win-Stay-Lose-Shift (WSLS) \cite{kraines_learning_1993} strategy that repeats the action from the previous round if the player is happy with the obtained payoff ($T$ and $R$), otherwise changes to the opposite action (when the payoff is $P$ or $S$). More strategies have been proposed and analyzed since then (see Martinez-Vaquero et al \cite{martinez-vaquero_generosity_2012} for a comprehensive study).

From an experimental angle, only a few works have examined how participant actions in the IPD may be translated into relevant strategies used by humans, while also studying what factors may affect the observed behavior. Some experiments with animals showed that guppies \cite{dugatkin_guppies_1988}, sticklebacks \cite{milinski_tit_1987} and tree swallows \cite{lombardo_mutual_1985}, exhibit a TFT-like behavior. In human behavioral economic experiments, both the theoretical GTFT and WSLS appear to align with the decision-process of the subjects \cite{milinski_working_1998}. Dal Bó et al \cite{dal_bo_strategy_2013} showed in another IPD experiment that, when given the choice among theoretical strategies, subjects preferred strategies like Always Defect (AllD), TFT, and Grim Trigger (starts cooperating, cooperates as long as both players have cooperated in the last round, and defects otherwise) and did not use WSLS, which was shown to have more "desirable" properties, such as not defecting forever after a deviation. 

While matching theoretical strategies to experiments or asking people to select their preferred strategy provides insight into how they relate to human preferences, it does not immediately reveal how humans actually decide to act in the IPD while playing freely, since their strategies can change with time and respond to different factors, such as their opponents' actions or simply by learning the game and reach their own equilibrium. Alternatively, their strategy may be directly inferred from the data itself, using algorithmic models \cite{breiman_statistical_2001}. This approach was taken to determine the strategies in an Ultimatum game experiment, where binary decision trees were used to model the strategies \cite{engle-warnick_inferring_2003} or symbolic regression \cite{duffy_using_2002}. A similar approach was taken to infer the behavior of participants in Trust games \cite{engle-warnick_inferring_2006, engle-warnick_evolution_2004} or in a market simulation, where strategies were inferred using using Bayesian inference \cite{engle-warnick_strategies_2002, kleiman-weiner_non-parametric_2018}. Yet so far, no inference of strategic models has been performed on IPD, a caveat this work is overcoming.

We present thus, on the one hand, the results of a behavioral economics experiment that investigates how subjects act in the IPD and, on the other hand, the strategies that can be extracted from the data. The experiment consists of two treatments: i) one where the subjects play the IPD with a fixed partner over a large, unspecified number of rounds, which will be called the \emph{fixed partners} or \emph{FP} treatment; and ii) a second long IPD treatment where the opponent changes each round, which we call \emph{shuffled partners} or \emph{SP} treatment. The objective of collecting the data over a large number of rounds was to understand what the effect is of a long experiment on the level of cooperation in the IPD and to study how the inferred behavior differs over time and between a setting where partnerships can be established or one where one is repeatedly confronted by strangers\cite{grujic_consistent_2012, andreoni_partners_2001, gachter_conditional_2007}. 

It is not clear from the literature how many rounds of the IPD are needed to observe a stabilization in the human decision-making process, meaning that the learning phase has passed and people are acting according to a clearly defined strategy. Some works have studied the strategies subjects play for ten rounds \cite{heuer_cooperation_2019}, another one that studied the evolution of cooperation in the IPD lasted on average 1.96 and 4.42 rounds, for their two treatments \cite{dal_bo_evolution_2011}, another work studied strategies in the IPD with noise lasted on average 8 rounds, where they noticed a considerable strategic diversity, suggesting the subjects did not learn the game completely \cite{fudenberg_slow_2012}, others, a range of 10 to 35 rounds \cite{fleis_once_2014}, 15 rounds \cite{majolo_human_2006} and 100 rounds\cite{liu_learning_2014}, to cite a few. Yet, the latter focuses on an interaction with a fixed theoretically defined agent. To study the evolution in the decision-making process, we need to know how long the subjects need to learn the game among themselves in different points in time. Given this insight, one can then use the algorithmic modeling approach to infer the actual behaviors, comparing them over time and over different treatment settings, as is the case here. 

As experience determines the behavior of each participant, which we will call context, we will use unsupervised clustering techniques to identify the contexts that are found in both IPD treatment data and then determine which strategic algorithmic models may be inferred from the data in each of the contexts (with the context being defined by their own actions and their opponents' in the previous round).  Our  hypothesis is that different behaviors will be inferred from the differing context experiences. Iterating the game for a sufficient time, as specified earlier, will thus allow to clearly discern their short and long-term behavior. We designed the experiment to last $100$ rounds, without explicitly informing this hard limit to the participants, which is much longer than previous experiments \cite{dal_bo_evolution_2011, fudenberg_slow_2012, fleis_once_2014} to cite a few, and investigated how many rounds are actually needed before the learning process ends and the participant behavior appears to become consistent. In other long experiments, this stabilization seemed to appear after 10-20 rounds\cite{grujic_three_2012}, which will be examined in more in a detailed analysis here. 

The participant behavior is modeled using Hidden Markov Models (HMM) \cite{rabiner_tutorial_1989}, and its parameters are trained using hmmlearn\cite{weiss_hmmlearn_2016} on the treatment data separated in behavioral clusters, while simultaneously trying to find the preferred minimal HMM structure (number of states and transition structure) to achieve this. The resulting HMM models are both simple and transparent, containing enough modeling power to represent subjects' strategies in the IPD while also being generative.  The latter is of interest as the inferred strategies could be directly be used within the context of theoretical simulations assessing their performance, which is left for future work.  

\section*{Methods} 

\subsection*{Experimental data collection}

As explained in the Introduction, data was collected for two treatments wherein participants played long IPD in two different pairwise configurations, i.e. fixed partners (FP) and shuffled partners (SP). The data from these experiments were collected in Brussels, Belgium, at the Brussels Experimental Economics Laboratory (BEEL), part of the Vrije Universiteit Brussel (VUB). Twelve sessions were held, in total 188 participants were recruited. The group of participants consisted of 48\% female and 52\% male. The average age in all sessions was 22$\pm$8 years old (see the Supplementary Information for the detailed data about the participants). All experiments followed the relevant guidelines and regulations of data protection and experiments with human participants and were approved by the Ethical Commission for Human Sciences at the VUB (ECHW2015\_3). Moreover, all participants gave consent to the experiment by signing a consent form after the instructions of the experiment were read and all questions the participants had were answered and addressed. The participants played the IPD on individual isolated laptops, designated to avoid any type of communication or arrangements by the participants, who remained anonymous throughout the experiment. At the end of the experiments, subjects were asked optional questions about their age, gender, and information about the game they just played (see the Supplementary Information for the detailed questions and answers).
 
 The FP treatment consisted of 6 sessions where 92 participants played pairwise IPD with the same opponent. The SP treatment consisted of 6 sessions where participants played the same IPD as FP but with a different opponent each round, hence the "shuffled" name. In the IPD one normally has a probability that the game is continued given by a parameter $\omega$, producing an average of $(1-\omega)^{-1}$ rounds. Also, in our design, participants were not told exactly when the IPD would end, and we designed them in such a way (see the descriptions in the Supplementary Information) that there were $100$ rounds. Table \ref{table:payoff} shows the payoff matrix containing the per round rewards used in the two treatments. For both treatments, participants could observe the actions of their partner in the previous round, even when this partner changed from the previous to the current round. 
 
\begin{table}[!ht]
\centering
\begin{tabular}{|c|l|l|}
\hline
\textbf{} & \textbf{C} & \textbf{D} \\\hline
\textbf{C} & 3,3 & 0,4 \\\hline
\textbf{D} & 4,0 & 1,1 \\\hline
\end{tabular}
\caption{Payoff matrix for both IPD treatments.}
\label{table:payoff}
\end{table}

At the beginning of the experiment, every participant read a detailed instruction document with a small test at the end to make sure they understood the game dynamics (see Supplementary Information). At the end of the experiment, participants were given their gains (mean = 7.85 euros, SD = 1.67 in the SP treatment and mean = 10.61, SD = 3.87 in the FP treatment) and a show-up fee of 2.5 euros. Most players spent a total of one hour from the moment they entered the computer room to the moment they were paid for their participation.

For each participant in both treatments we collected their choices, as is visualized by panel A in Supplementary Figure \ref*{fig:hmm_example}. The combination of two actions, i.e. $CC$, $CD$, $DC$ or $DD$ (action format: player-opponent, e.g. $CD$ means the focal player cooperated and their opponent defected in the previous round), provides a context for the next round, as participants will get this information when making their decision.  Each context can now be combined with the action of a participant after observing that context and this combination, as shown in panel B of Supplementary Figure \ref*{fig:hmm_example}, can now be translated into one of eight values. The entire sequence of actions of a participant and the associated contexts can thus be transformed into a new sequence of numbers that represent the \emph{conditional actions} of each participant.  This new sequence will be used to train an HMM, representing in its emission probabilities in every state the conditional response, i.e. cooperate or defect, based on the context they experienced.  We also collect for each context, the probability of observing a conditionally cooperative action, as visualized in Figures \ref{fig:prob_coop_fix} and \ref{fig:prob_coop_rand} in the Results section.

\subsection*{Clustering contexts and context-dependent behaviors}

\begin{figure}[!ht]
\centering
 \includegraphics[width=0.90\linewidth]{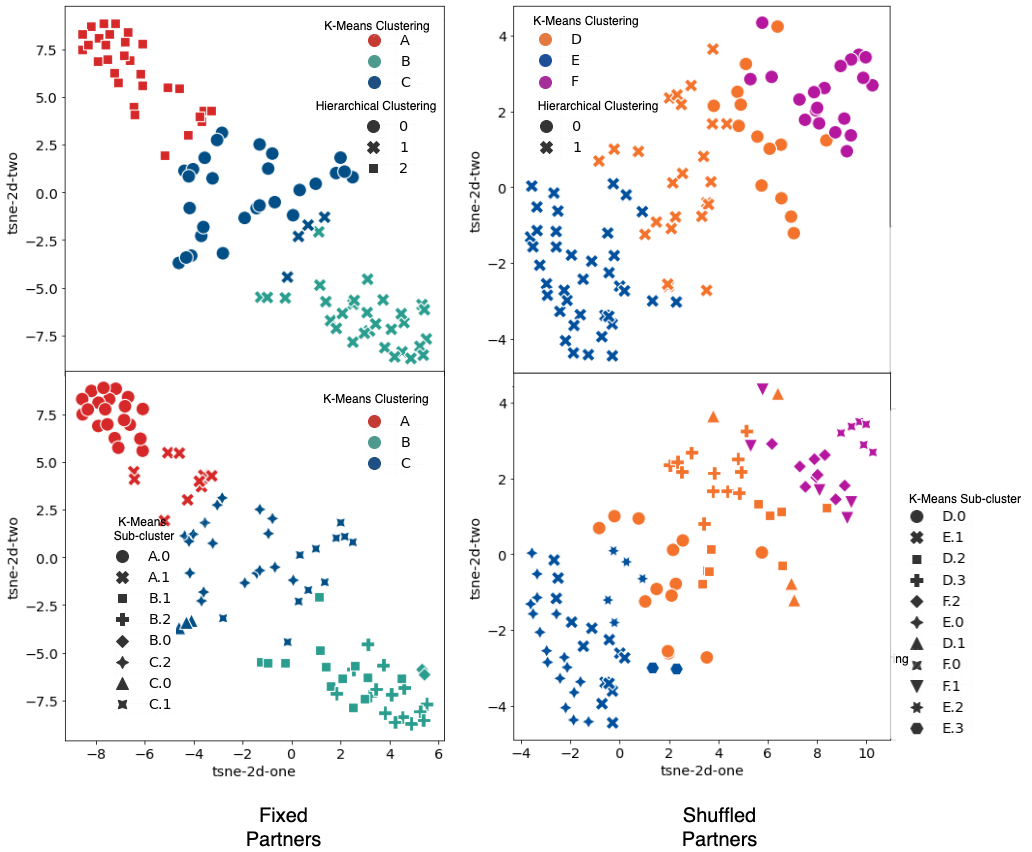}
\caption{tSNE components of both treatments reduced to two components. The tSNE plot was made with the context variables (CC, CD, DC and DD) and the cooperation variables (CCC, CDC, DCC, DDC), i.e. the number of times each subject faced each one of these scenarios in the previous round, for example, CD means that the focal player cooperated and their opponent defected in the previous round. In this case, these eight variables were reduced to two main components for visualization, each point represents a subject. In the top row, each color represents the classification in the K-Means clustering and its shape the hierarchical clustering method for comparison. The left panel in the top row shows the subjects of the FP treatment (clusters A, B and C of K-Means) and the right panel the SP treatment (clusters D, E and F of K-Means). The third row shows the behavioral clusterings. It is visible how the clustering made with different methods overlap in both treatments. In the row below, it is shown how the clusters in k-means are subdivided by their corresponding sub-clusters. }
\label{fig:tsne}
\end{figure}

People act according to their preferences, which includes how they think others should behave. In order to understand the strategies that are being used one needs to explicitly consider the context wherein they happen. Thus, to find the strategies in the FP and SP treatments, we first separate participants according to their experiences. Once this is complete one can question whether each participant displays the same response for the contexts they experience. This allows one to correctly grasp how humans act in the IPD. To identify the context groups and the behaviors within those groups, a cluster analysis was performed here:  we first clustered the subjects based on the number of times of $(CC)$, $(CD)$, $(DC)$ and $(DD)$ happened, providing a \emph{contextual clustering}. Subsequently, another clustering was performed on the number of cooperative actions in the sequence for each of the previous action combinations, i.e. $(CC)C$, $(CD)C$, $(DC)C$ and $(DD)C$, providing thus a \emph{behavioral clustering per context}. These variables are used to generate the t-distributed stochastic neighbor embedding (t-SNE)\cite{maaten_visualizing_2008} plot in Figure \ref{fig:context_cluster_chart} and Supplementary Figure \ref*{fig:tsne}.

\begin{figure}[!ht]
\centering
 \includegraphics[width=\linewidth]{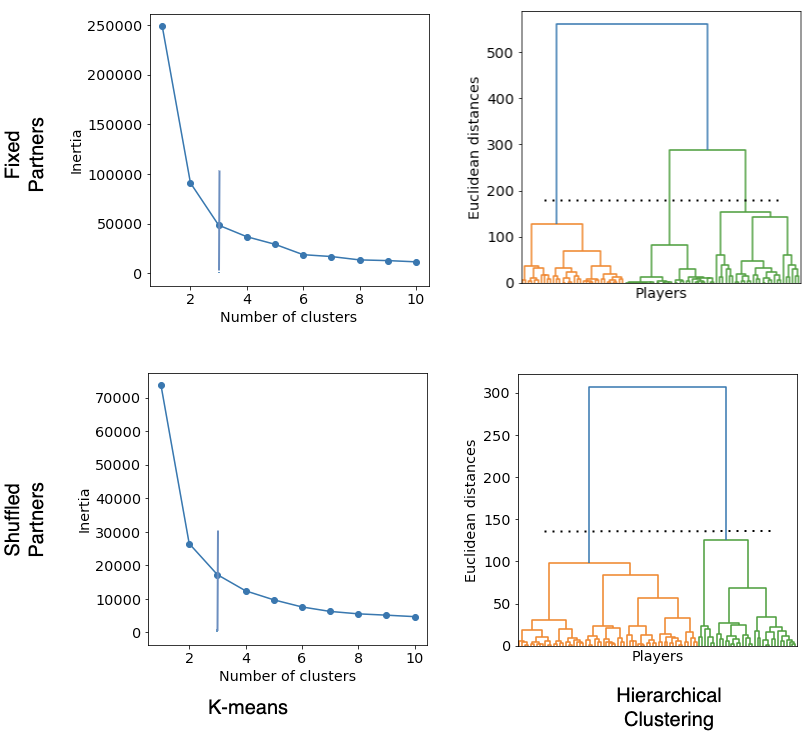}
\caption{Number of clusters per treatment. To test the number of clusters, in K-means (left panel) uses the "elbow" method, where a range of different $k$ (x-axis) are tested for their sum of squared distance between each data-point to its centroid, this is what is called inertia (y-axis). The vertical line represents the "elbow" of the curve, as explained in the Methods section, or the point where the curve starts to stabilize. We used the Python library kneedle by Satopaa et al. \cite{satopaa_finding_2011} to find this point. Hierarchical clustering (right panel) uses a dendrogram, where all data points are plotted according to their distance to each other. To pick the number of clusters, one follows the largest vertical distance from the top and crosses an horizontal line (dotted line). Fewer clusters were preferred at this stage, for example picking 5 clusters in FP resulted in $0.5571$ and picking 3 yielded $0.5531$. For both K-means and Hierarchical Clustering, we used the scikit library for Python \cite{pedregosa_scikit-learn_2011}.}
\label{fig:curves_cluster_context}
\end{figure}

Different clustering approaches were considered (e.g. K-means, Hierarchical Clustering, and Network Modularity analysis, as can be observed in Supplementary Information, see section \ref*{methods:mod_net} for more information on the Modularity Network clustering) but in the end, K-means is sufficient as the other approaches generated similar results (see for Supplementary Figure \ref*{fig:curves_cluster_context}, Figure \ref*{fig:tsne} top row, for a comparison with Hierarchical Clustering and Figure \ref{fig:modnet} for Modularity Network Clustering). The implementations were done in Scikit learn \cite{pedregosa_scikit-learn_2011}. To determine the optimal number of clusters, we used the "elbow" method by Santopaa et al \cite{satopaa_finding_2011} which plots a curve with the sum of squared distances of samples to their closest cluster center and chooses the minimum best number of clusters that minimize the inertia. See Supplementary Figures \ref*{fig:curves_cluster_context} and \ref*{fig:curves_sub_cluster} for the results produced but the "elbow" method. 

To evaluate the quality of the results generated by clustering algorithms, we use the silhouette measure\cite{rousseeuw_silhouettes_1987}, which measures the mean intra-cluster Euclidean distance and the mean nearest-cluster Euclidean distance for each observation, with 1 is being the best score and values near zero indicate that some clusters might be overlapping, while negative values indicate observations assigned to a wrong cluster.

\subsection*{Inferring Hidden Markov Models}

\begin{figure}
\centering
 \includegraphics[width=0.8\linewidth]{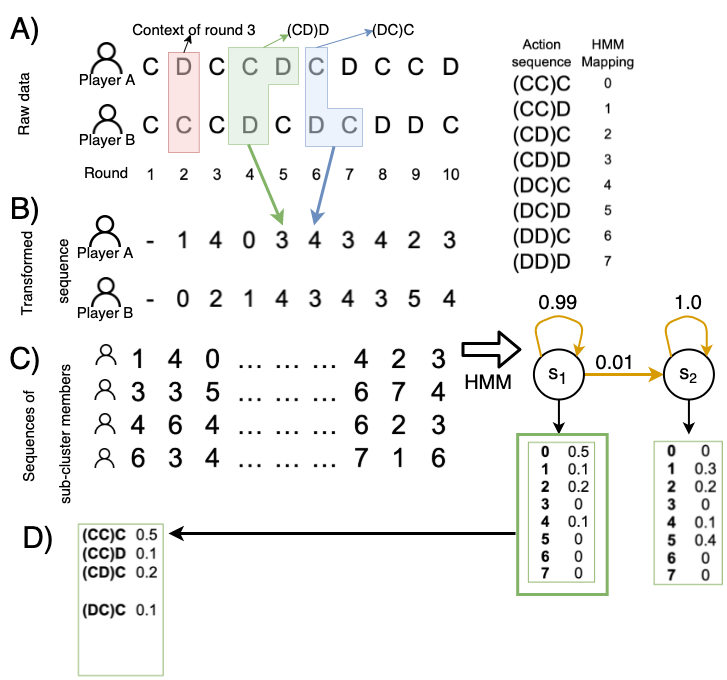}
\caption{From treatment data to HMM. \textbf{A)} At the top, the first 10 rounds between a player and her opponent, In the red box, an example of the context for round 3 The green box represents the action sequence $(CD)D$ for player A and the blue $(DC)C$ for player B in rounds 5 and 7 respectively. \textbf{B)} The transformed sequence for players A and B into integers so the HMM with the mapping table at the right. \textbf{C)} At the bottom, the resulting sequences are grouped given their sub-cluster and then processed by the HMM. In this example, the resulting HMM is a model with two states $s_1$ and $s_2$ with transition probabilities given by the yellow arrows with emission probabilities in the boxes below. \textbf{D)} HMM visualisation. Note that all the emission probabilities $<= 0.05$ and all transition probabilities $<=0.01$ were not taken into account for readability. The resulting HMM emission table is coloured green from the hidden state $s_1$.}
\label{fig:hmm_example}
\end{figure}

Given the context clusters and the context-dependent behavioral sub-clusters, a HMM, using the conditional action sequences (see Figure \ref*{fig:hmm_example}), is produced for each sub-cluster. As shown in the figure, a HMM is composed of a number of states ($s_1$ and $s_2$ in the figure) which are connected by transitions (yellow arrows). The HMM is a probabilistic Markov Model in which each observation of a sequence is produced by a hidden (non-observable) state \cite{rabiner_tutorial_1989}. We used a multinomial model for HMM with a sequential structure, this means that the model has states connected from left to right, where a transition to another state can only be made in that direction. Returns to a previous hidden state are thus not possible. As also shown in the figure, the transformed sequence of conditional actions is used to  train the HMM.

The procedure to determine the optimal number of hidden states is specified in Algorithm \ref{alg:hmm}. Up to 4 hidden states were tested, and the selected number of states $h$ corresponded to the number of different hidden states predicted by the model itself, a low number of hidden states is preferred, both to avoid overfitting and to facilitate interpretation. One last reason to pick fewer hidden states is that some of these states and their transition probabilities are small, which between runs can tend to zero, and it will end up leaving the hidden state disconnected. To ensure connectedness, no model was accepted with a probability of transition less or equal to 0.01. These models were fitted over one million trials and picked the one that resulted with the highest log-likelihood to fit the sequences.

\begin{algorithm} \caption{Hidden Markov Model per sub-cluster}\label{alg:hmm}
\begin{algorithmic}[1]

\ForEach {sub-cluster $s \in $ cluster $c$ }
\State $X_{train}$, $X_{test}$ = data[sub-cluster == s]
\State ll\_max = 0, best\_model = null
\ForEach {$h \in $ [4,3,2,1] \Comment{Number of Hidden States} }

\State model = MultinomialHMM(h)
\State trans\_matrix, emission\_matrix, initial\_probs = model.fit($X_{train}$)
\State ll = log\_likelihood(model.predict($X_{test}$))
\State if ll $>$ ll\_max AND $p \in $ trans\_matrix $>= 0.01$: best\_model = model
\EndFor
\EndFor
\end{algorithmic}
\end{algorithm}

 To train the HMM, the hmmlearn library\cite{weiss_hmmlearn_2016} was used. To visualize the resulting models the GraphViz package for Python\cite{bank_graphviz_2021} was used. We expect the participant choices to be stochastic since subjects were not instructed to use any strategy in particular, but the action they considered was best according to their expectations. For this reason, we expect the inferred strategies to be noisy. For visualization, we mapped the number sequences back to human-readable triplets, and all the emission probabilities in the HMM less than $0.05$ were discarded, as shown in Figure \ref*{fig:hmm_example}.

\subsection*{Evolution of the strategies}

To analyze how the players changed and adapted their strategy over all the $100$ rounds, the same procedure of the two-fold clustering and HMM modeling was performed on four different round windows, i.e. from round 1-25, 26-50, 51-75, and finally rounds 76-100). The objective is to assess whether the strategy stabilizes over time and to examine how the treatment, i.e. fixed or shuffled partners, affects the results.

\section*{Results}

\subsection*{Specific contextual subgroups with associated behavioral responses emerge in each treatment}

\begin{figure}[ht!]
\centering
 \includegraphics[width=0.7\linewidth]{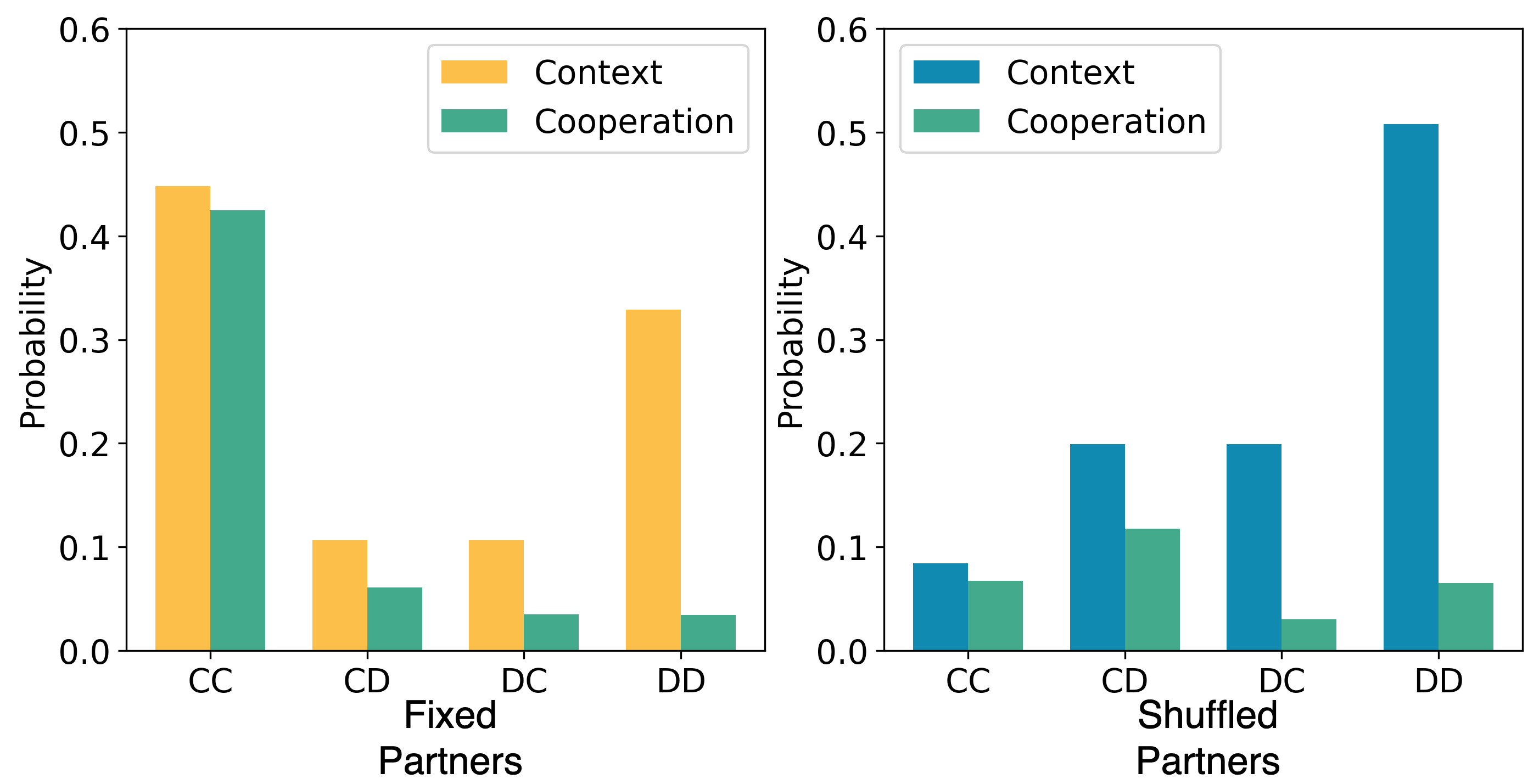}
\caption{Probability of each context and the cooperation rate per treatment. In the x-axis, the context of each decision is shown, i.e. the actions of the previous round. The green bar in each chart shows the fraction of  subsequent cooperative actions when experiencing the particular context. This way, in FP (panel A), there are many rounds with context CC (mutual cooperation) and DD(mutual defection). In SP (panel B)  DD occurs clearly more often.}
\label{fig:context_chart}
\end{figure}

Before identifying the different clusters in the data provided by both treatments (see Methods and Supplementary Information for details), we first examine the distribution of contexts  experienced by participants during the experiment. Figure~\ref{fig:context_chart} shows this distribution while also revealing the fraction of times each context led to a cooperative response by a participant. In the case of the FP treatment, mutual cooperation ($CC$) or mutual defection ($DD$) is observed most often and was matched with the expected response (i.e. either $(CC) \rightarrow C$ or $(DD)\rightarrow D$, which we write as $(CC)C$ and $(DD)D$ respectively). In the SP treatment, mutual defection $DD$ and the anticipated response $D$ was most prevalent. Although a similar cooperation probability was shared by the two treatments (see Supplementary Table~\ref*{table:context}) in the case of $DC$ and $CD$, there was a higher probability of cooperation in FP than in SP in the $DC$ context, i.e. responding positively to an act of cooperation of the co-player. Overall, the figure shows that by shuffling partners cooperation is reduced, as anticipated by prior experiments \cite{grujic_consistent_2012}. Also, it already hints at different clusters that could be inferred from both data sets.

Clustering the treatments on the contextual information that each participant experiences (the frequency of contexts $CC$, $CD$, $DC$ and $DD$) with K-means (see Methods) singled out 3 groups in both FP and also 3 in SP. As reported in Methods and visualized in Supplementary Figure \ref*{fig:curves_cluster_context}, different algorithms (network modularity and hierarchical clustering) were tested, revealing that a similar number of clusters was obtained in each case. Cluster quality is assessed using silhouette scores, which are $0.5445$ and $0.4201$ respectively for the FP and SP treatment. These results indicated that a relatively good separation is found between the different contextual clusters.

\begin{figure}[!ht]
\centering
 \includegraphics[width=\linewidth]{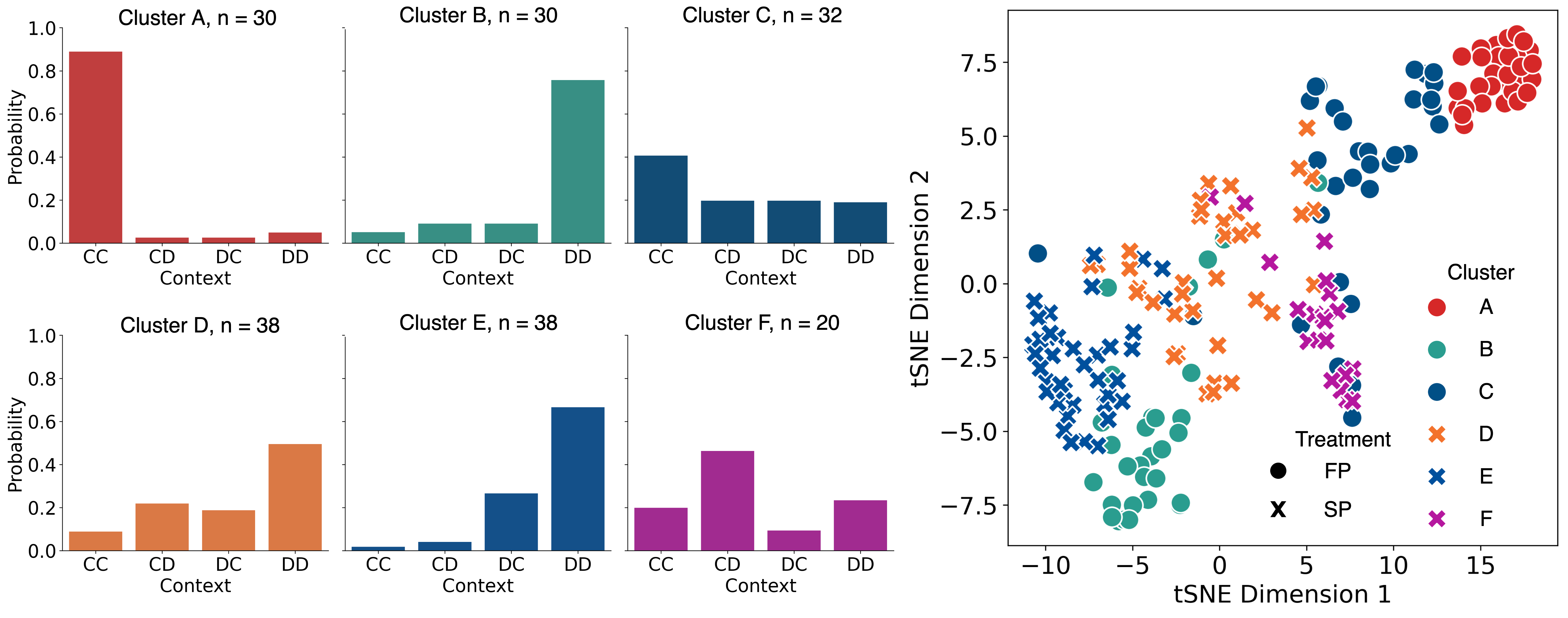}
\caption{Probability of cooperation given each context per cluster (using K-Means). \textbf{A)} Top row shows the context composition of each cluster in the FP treatment, the bottom row does the same for each SP cluster. In the FP treatment, clusters A and B are composed of opposing experiences, either $CC$ for A or $DD$ for B. Cluster C is a mix with a bias towards mutually cooperative encounters. In the SP treatment, bottom row, the experiences tend to be more of the $DD$ kind, as shown in clusters D and E. Cluster F in SP showed a mixed experience, yet with a different context composition as  C. \textbf{B)} tSNE visualization of the different clusters A-F. As can be seen, cluster A and B lie at the two extremes of this embedding. The clusters D and E have more in common with the defective behaviors in B than the cooperative ones, i.e. A and C. C differentiates itself from the defective clusters as  $CC$ is more likely to occur, yet part of C overlaps with the behaviors observed in F.}
\label{fig:context_cluster_chart}
\end{figure}

Panel A in Figure \ref{fig:context_cluster_chart} shows the composition of each cluster in terms of its context. In the FP treatment (A, top row), three different groups are identified: Cluster A containing the players who experience mutual cooperation $CC$ most often, cluster B where they experience mutual defection $DD$ most, and then cluster C where experiences are mixed. These results were anticipated, given the differences shown in Figure \ref{fig:context_chart}. In SP (bottom row Figure \ref{fig:context_cluster_chart}A) mutual defection is favored over mutual cooperation in two of the clusters, i.e. clusters D and E. Nonetheless, there are sufficient differences between them which are captured by our clustering approach. 

The tSNE plot (see Methods) in panel B of Figure \ref{fig:context_cluster_chart} allows for the exploration of these differences and compares the clusters identified in both treatments. This plot reveals that clusters A and C, which are more cooperative in experienced contexts and responses, clearly differentiate themselves from the rest, and one can find them at completely the other side of experiences and responses to those belonging to cluster B of the FP treatment. There seems nonetheless to be an overlap between some members cluster C and those in F in the SP treatment, yet most C members form a group on themselves. The experiences and behaviors of members of the E cluster, on the other hand, are close to those in B, which one can also observe when comparing the two bar charts. Finally, cluster D consists of participants that have experiences and responses that lie in between all others, separating the more defective and cooperative spectrum. 

Given these observations, in the following sections, we show how we can infer the behaviors adopted by participants in each contextual cluster. It is important to note that not considering explicitly the experiences of the participants and focusing directly on the conditional behavior may result in grouping together participants that have encountered a different distribution of contexts.  It is therefore essential to first partition the participants in function of their contexts distribution and then determine how they differ in their choices given these experiences.

\begin{table}[!ht]
\centering
\begin{tabular}{|c|c|c|c|c|}
\hline
\textbf{Treatment} & \textbf{Context } & \textbf{No. behavioral} & \textbf{Silhouette}\\
& \textbf{Cluster} & \textbf{sub-clusters} & \textbf{ score}\\
\hline
 FP & A ($n=30$) & 2  & 0.6868 \\ 
  FP & B ($n=30$) & 3  & 0.3848 \\ 
  FP & C ($n=32$) & 3  & 0.4627 \\ 
  \hline
  SP & D ($n=38$) & 4  & 0.2993 \\ 
  SP & E ($n=38$) & 4  & 0.4417 \\ 
  SP & F ($n=20$) & 3  & 0.4176 \\ 

 \hline

\hline
\end{tabular}
\caption{Summary of the clustering results using K-Means . The contextual clustering results in 3 clusters of similar sizes. The behavioral  clustering per context yielded 8 behavioral sub-clusters in FP and 11 behavioral sub-clusters in SP. The scikit-learn \cite{pedregosa_scikit-learn_2011} silhouette analysis was used to determine the quality of the clusters generated with the "elbow" method (see Methods).}

\label{table:sub-clustering}
\end{table}

\subsection*{Fixed partners promotes behavioral self-selection to cooperative or defecting behaviors}

\begin{figure}[!ht]
\centering
  \includegraphics[width=\linewidth]{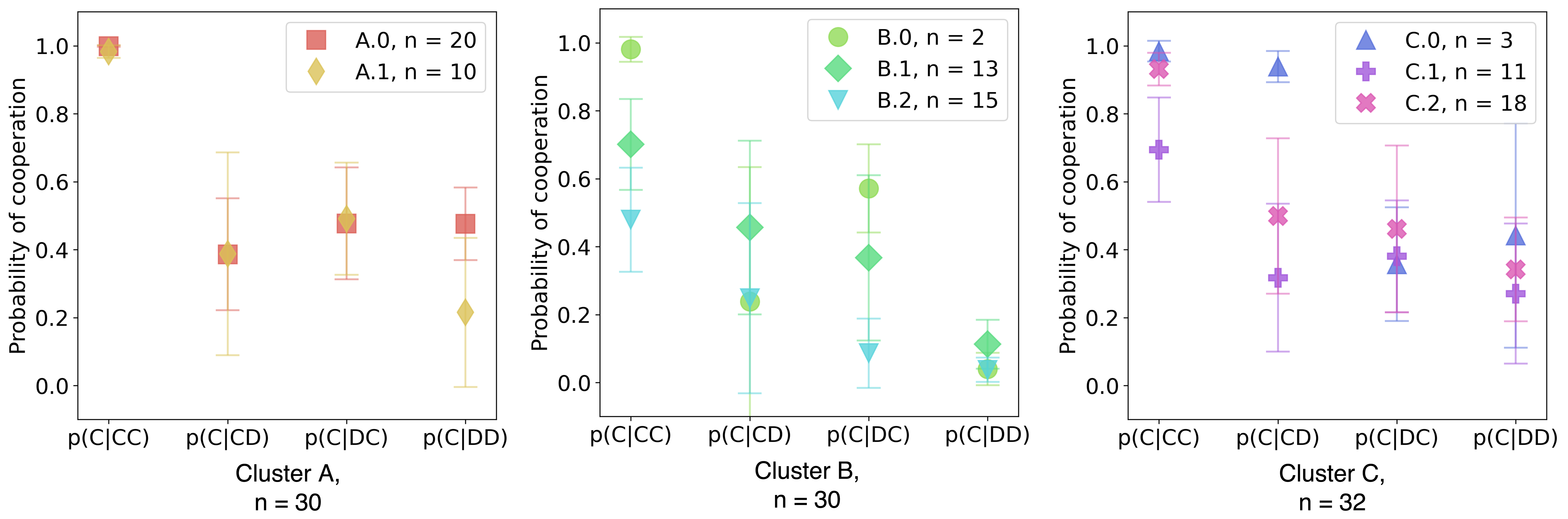}
\label{fig:prob_coop}
\caption{Probability of cooperation given each context per cluster for the FP treatment. In each plot, the conditional probability of cooperating for each context is given. The bars represent the binomial error. Each contextual cluster is divided into sub-clusters that were identified using the frequency of cooperation given a context. It can be seen that in cluster A (left) that the differences between the sub-clusters are due to the difference in cooperation for the context $DD$. In cluster B (center), they all have a similar response to the $DD$ context, which explains the distribution seen in Figure \ref{fig:context_cluster_chart}. Cluster C (right) shows a mixed experience among the sub-clusters, with cluster C.0 showing significantly stronger cooperative response in case of the context $CD$.}
\label{fig:prob_coop_fix}
\end{figure}

\begin{figure}[!ht]
\centering
\includegraphics[width=\linewidth]{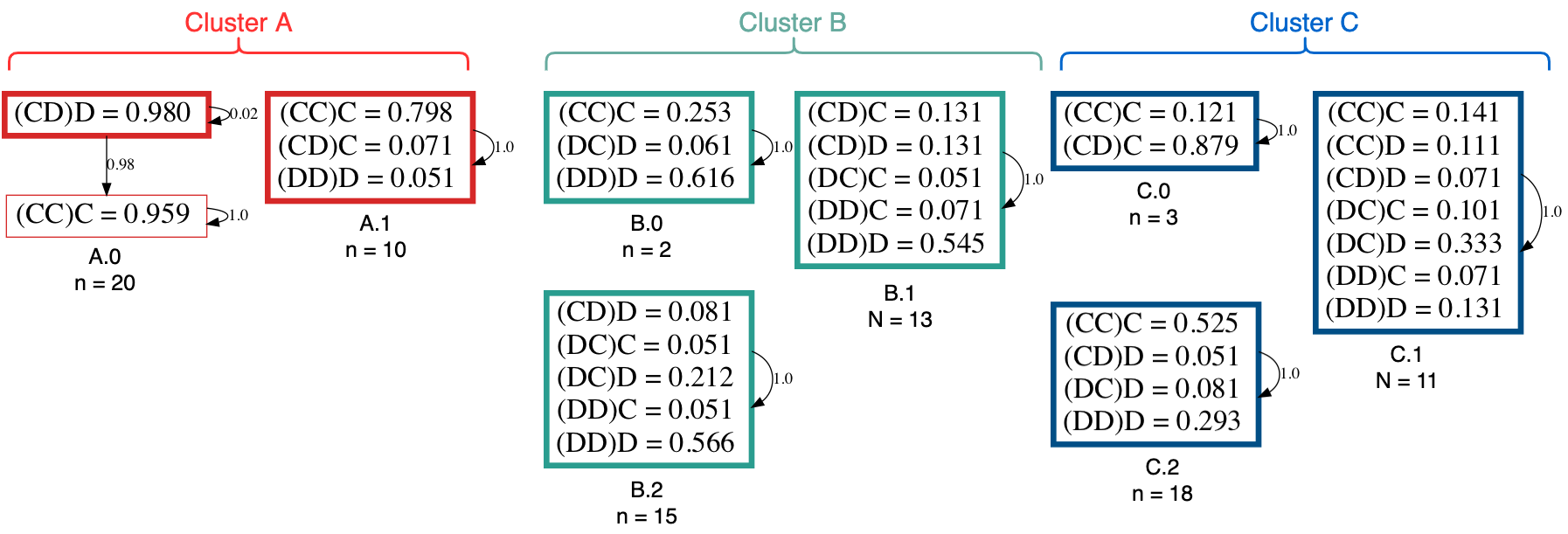}
\caption{Hidden Markov Models for the FP treatment. Here, the eight sub-clusters found in FP have a HMM that describes each sub-cluster's strategy for the data over all rounds. Bold rectangles represent the initial state, while the others represent subsequent hidden states. Symbols with a probability lower than 0.05 are not shown, as well as the transition probability between states lower or equal to 0.01. For example, in sub-cluster A.0, the subjects play $D$ in the context $CD$ with a very small probability to continue, but a big one to pass to the other hidden state in the sequence, where they mutually cooperated ($(CC)C$). On the other hand, sub-cluster A.1 has some probability to defect given mutual defection $(DD)D$, which A.0 does not show. Overall, it appears a model with only one state is preferred by the training method (see Methods).}
\label{fig:strat_FP}
\end{figure}

\begin{figure}[!ht]
\centering
 \includegraphics[width=\linewidth]{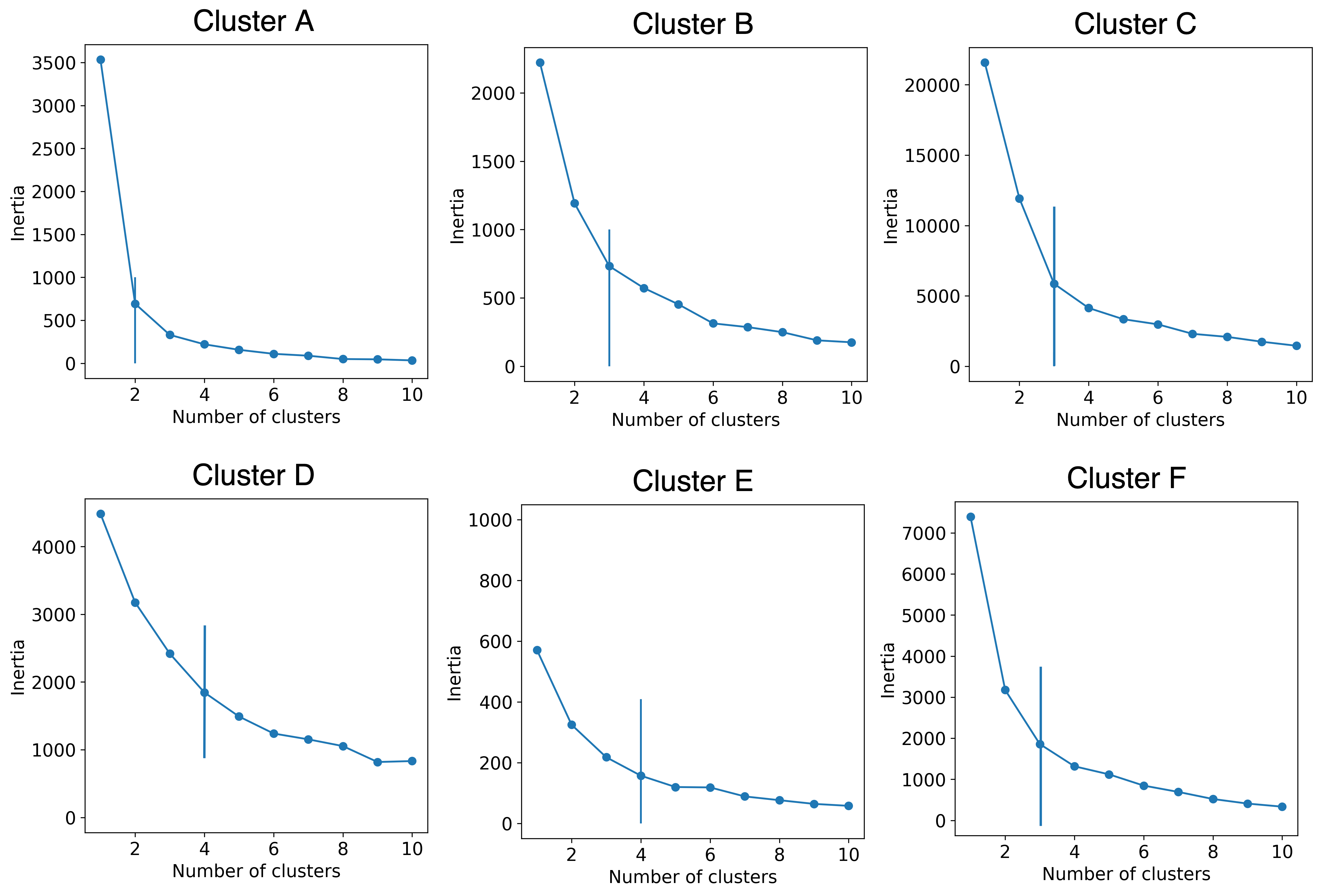}
\caption{Number of sub-clusters per cluster. A curve of inertia is plotted per cluster and per treatment. Inertia is the sum of squared distances of each data-point to their corresponding centroid in K-means (y-axis), and a range of different $k$ are tested (x-axis). The vertical line represents the "elbow" of the curve, as explained in the Methods section, or the point where the curve starts to stabilize. We used the Python library kneedle by Satopaa et al. \cite{satopaa_finding_2011} to find this point. In the top row, the FP is shown with its clusters: A, B, C; and in the bottom row the SP treatment with its clusters: D, E and F. }
\label{fig:curves_sub_cluster}
\end{figure}

Table~\ref{table:sub-clustering} reports the number of behavioral sub-clusters for each contextual cluster (see Methods): in total, 8 subgroups were inferred from the raw data of the FP treatment, i.e., 2 in cluster A, 3 in cluster B and 3 in cluster C. Supplementary Figure~\ref*{fig:tsne} shows a tSNE visualization of how the treatment data is divided into clusters and sub-clusters for FP. As before, the best settings for $K$ in K-means clustering was produced using the "elbow" method \cite{satopaa_finding_2011} (see Methods and  Supplementary Figure \ref{fig:curves_sub_cluster}). 

The behaviors in each contextual cluster are captured each time by two plots:  i) a first plot that shows the likelihood of cooperation when experiencing a given context in a contextual cluster, and ii) a second plot that provides the inferred HMM in the behavioral sub-cluster (see Methods). Figures \ref{fig:prob_coop_fix} and \ref{fig:strat_FP} provide both plots for all the sub-clusters in each contextual cluster for the FP treatment. 

For cluster A, one can observe the difference in response when it comes to a $DD$ situation, when comparing both sub-clusters (results in red):  Individuals in cluster A.0 are more likely to cooperate again, than those in A.1 (see Figure~\ref{fig:prob_coop_fix}). Moreover, the resulting HMM for cluster A.0 shows how an initial strategy of reciprocating defection ($(CD)D$), leads to mutual cooperation ($(CC)C$). In addition, an increase in forgiving defective behavior ($(CD)C$) can be observed in Cluster A.1 (see HMM for A.1 in Figure~\ref{fig:strat_FP}), which is present less often in cluster A.0. Note that for this and following HMM, more elaborate models (with more states and transitions) could be constructed. As explained in Methods, we identify the optimal number of states for the model so as to avoid overfitting and to keep the HMM easy to understand. 

Although cluster B is divided into three sub-clusters, there are actually two essential ones in terms of the number of participants they represent (see both Figure~\ref{fig:prob_coop_fix} and the green HMM in Figure~\ref{fig:strat_FP}). This next FP cluster mainly experienced mutual defection, and it is situated at the opposite of the behaviors present in the other two FP clusters (see Figure \ref{fig:context_cluster_chart}B). Looking at B.1 and B.2 in both figures, participants in sub-cluster B.1 appear to respond more often with cooperation than those in sub-cluster B.2 (see Figure \ref{fig:prob_coop_fix}, center panel), given that the triplets $(CD)C$ and $(DD)C$ occur at a higher frequency and that unconditional defection $(DC)D$ occurs with a high frequency in sub-cluster B.2. The two participants within cluster B.0 appear to be outliers with respect to the other members of the cluster as they were able to establish mutual cooperation more often than the participants in both other sub-clusters in cluster B. 

Finally, in cluster C of the FP treatment, participants experienced a mixture of contexts, with a preference for mutually cooperative interactions. Clustering this group and inferring the corresponding HMM still revealed some differences (see again Figure~\ref{fig:prob_coop_fix}, Cluster C and Figure~\ref{fig:strat_FP}, blue group). Sub-cluster C.0 presents different behavior, members of this sub-cluster unconditionally cooperated despite their opponent's defection ($(CD)C$). Sub-cluster C.1 had a mixed strategy of reciprocating their opponent's previous action, but their rate of exploitation ($(CC)D$ = 0.33 and $(DC)D$ = 0.07) is higher than other sub-clusters. Lastly, the sub-cluster C.2 was mainly matching their opponent's previous action.

So clusters A and B in the FP treatment consist mostly of (forgiving as well as strict) cooperators and defectors respectively. It appears to indicate that a behavioral self-selection occurred among the participants in FP. In complex systems and evolutionary game theory, a self-selection mechanism (also called self-organization) occurs when parts of the system appear to reach a stable state \cite{yackinous_chapter_2015}. In this case, self-organization brings them to either of the extreme cases, i.e. mostly cooperation or mostly defection. This phenomenon could be happening because of the possibility for imitation of the choices by their neighbors in FP, as argued by Mahmoodi et al. \cite{mahmoodi_self-organizing_2017}. Members of the third cluster are in some sense still in between, either switching between matching choices or trying to outsmart the co-player.

\subsection*{While stranger interactions induce defection, cooperative aspirations remain}

\begin{figure}[!ht]
\centering
  \includegraphics[width=\linewidth]{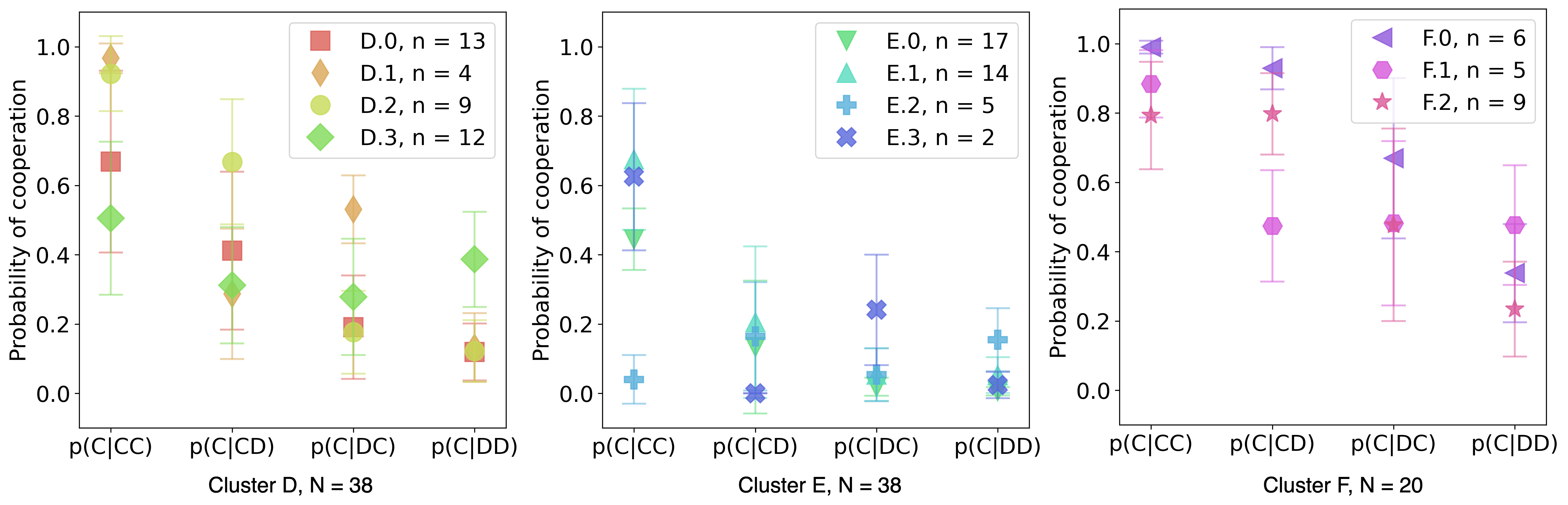}
\caption{Probability of cooperation given each context for the Shuffled Partners treatment. In the x-axis, it is shown the probability of cooperating given that in the previous round there was a certain context. The bars represent the binomial error. Each cluster is divided into sub-cluster that were divided using the frequency of cooperation given a context. As in FP (Figure \ref{fig:prob_coop_fix}), the sub-clusters present differences even when in their clusters a certain context dominates. For example, in p($C|CC$), it is shown how sub-cluster E.2 has a very low probability to cooperate given cooperation in the previous round, while the other sub-clusters are situated between 0.4 and 0.6. }
\label{fig:prob_coop_rand}
\end{figure}

\begin{figure}[!ht]
\centering
\includegraphics[width=\linewidth]{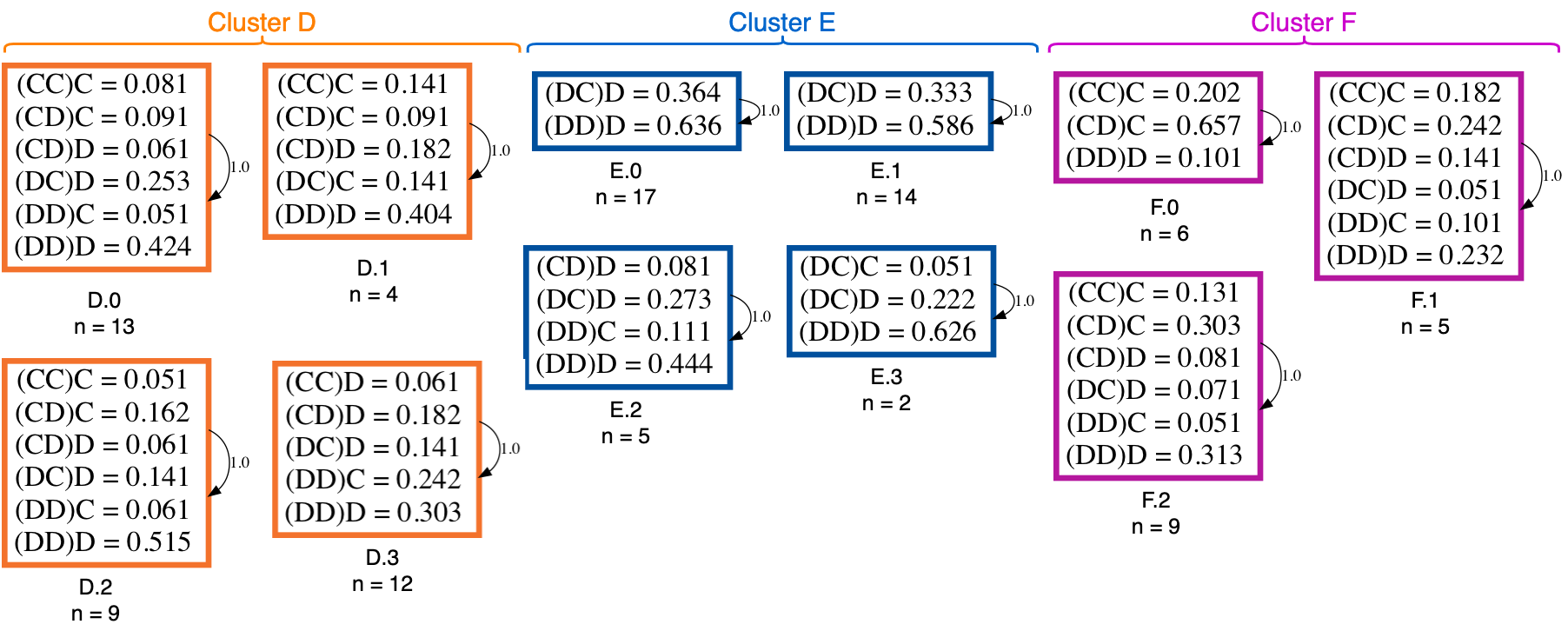}
\caption{Hidden Markov Models for the Shuffled Partners treatment. Here, the eleven sub-clusters found in SP have a HMM that describes each sub-cluster's strategy. Symbols with a probability less than 0.05 were not shown, as well as the transition between states less or equal to 0.01. Then, for example in cluster E.1, it is shown how the subjects here played unconditionally defect with a probability of $0.919$, the other symbols' probabilities ($(CC)C$, $(CC)D$, etc.) are not shown.}
\label{fig:strat_SP}
\end{figure}

In the SP treatment, a larger variety of  behaviors can be observed, i.e. 11 in total spread again over 3 contextual clusters. Figures \ref{fig:prob_coop_rand} and \ref{fig:strat_SP} immediately reveal some interesting information about the participants' behavior in the three clusters found for the SP treatment. Over all three clusters, i.e, D, E and F, we can observe, through their associated HMM models, that they have different degrees of mutual defection ($DD$), with E containing the most (as also was clear from Figure \ref{fig:context_cluster_chart}, bottom row, panel A). Moreover, the HMM models for E reveal a high tendency to defect even when the co-player acts cooperatively, even more than the defecting cluster B in FP. The opposite seems to occur in cluster F, where mutual cooperation is the highest and players are more likely to ignore a unilateral defection of the co-player. These two SP clusters thus represent opposing behavior given by their differences in context, as can also be seen in the tSNE plot in Figure \ref{fig:context_cluster_chart}B: clusters E and F are most distant in the SP treatment data in the tSNE embedding (More details can be observed in Supplementary Figure~\ref*{fig:tsne}). Based on Figure~\ref{fig:context_cluster_chart}B we also saw that F has a strong overlap with  a subset of cluster C of the FP treatment, i.e. the group with a mixed contextual distribution, biased towards mutual cooperation. Considering the HMM, this mapping appears to be the closest to the sub-cluster C.0 between and F.0.  This overlap is most likely to the high likelihood of continuing cooperation even when the co-player defected, i.e. $(CD)C$. For the rest, there appears to be little overlap between the behaviors in both treatments, demonstrating that human responses in both treatments are apparently different.

Cluster D on the other hand, appears to represent intermediate behavior with players still more mutually defecting but the frequency of some other response patterns has increased, which is also visible in Figure \ref{fig:context_cluster_chart}B.  Sub-cluster D.1 shows to be more conditionally cooperative ($(CC)C$ and $(DC)C$) than the other sub-clusters in D, i.e. with probability 0.3 (see Figure \ref{fig:strat_SP}, cluster D). Players in sub-cluster D.3 and E.2 appear to try to signal their co-players to cooperate given the high frequency of $(DD)C$ in that subgroup to establish cooperation. The cluster D.0 is very similar to the sub-cluster D.2, where the difference is how forgiving and relentless they appear to be in relation to their co-player (see frequency of $(CD)C$ and $(DC)D$ in D.0 versus D.2). These variations between the HMM models are highly interesting as they reveal that although some general rules may be inferred on the behaviors, there are some important variations in the responses that push the overall outcome of the interactions towards cooperation or defection. 

\subsection*{Decision-making simplifies over time and outcomes are decided early}

\begin{figure}[!ht]
\centering
\includegraphics[width=0.8\linewidth]{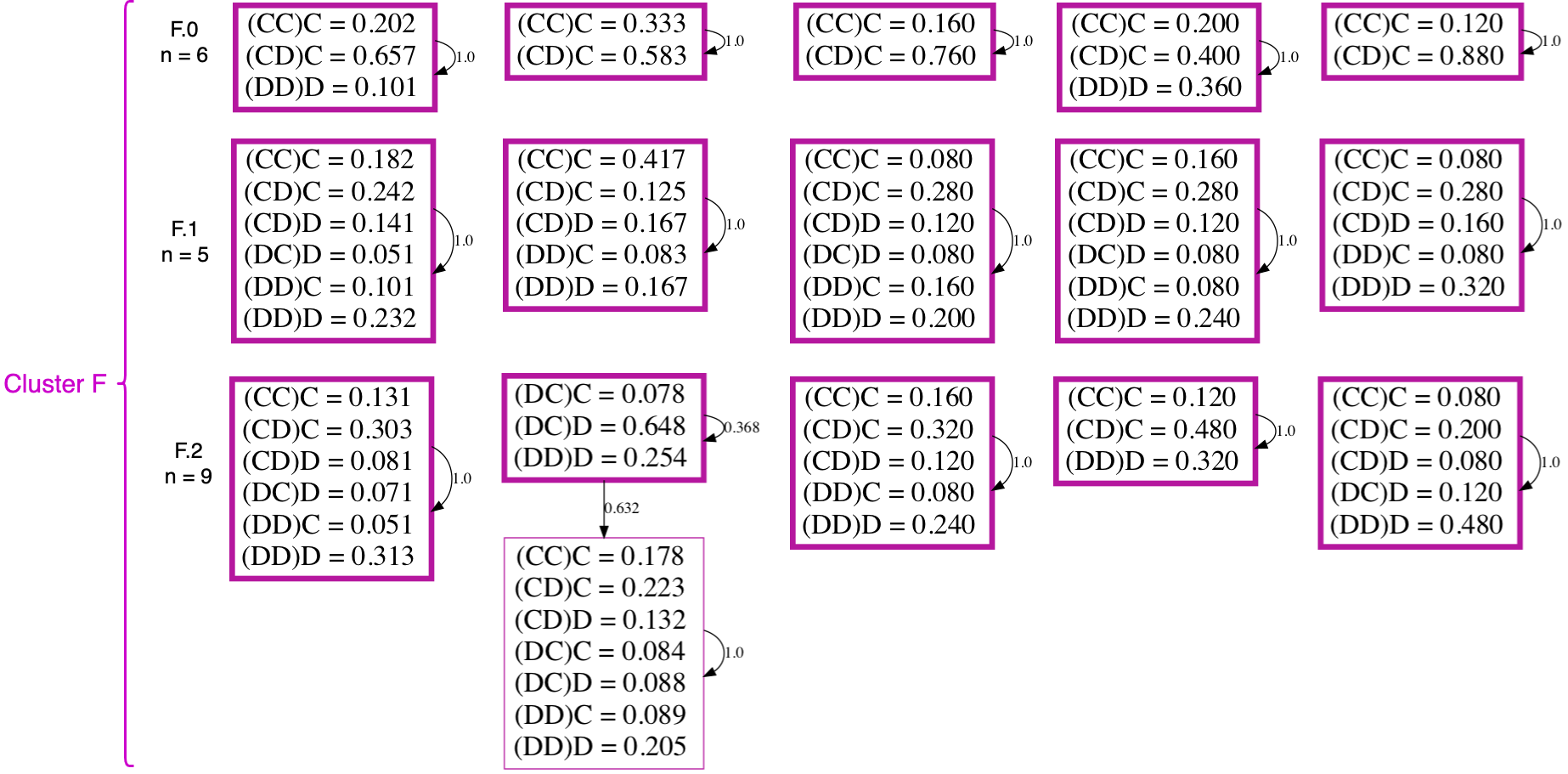}
\caption{Hidden Markov Models for the Shuffled Partners treatment. Here, the ten sub-clusters found in SP have a HMM that describes each sub-cluster's strategy. Bold rectangles represent the initial state, while the others represent subsequent hidden states. At the top, round windows are specified, starting from 1-100, then in windows of 25.}
\label{fig:strategies_SP_rounds}
\end{figure}

Although the HMM models in Figures \ref{fig:strat_FP} and \ref{fig:strat_SP} provide detailed insights into the behaviors observed over a complete (long) IPD experiment, we need to understand whether these behaviors are consistent over time, or whether early decision-making differs from downstream decision-making, which would indicate a learning effect in the experiment. To achieve this goal, the treatment data for each participant is divided  into four parts, each consisting of 25 rounds. For each window of 25 rounds, a separate HMM is trained, collectively visualized in the Supplementary Figure \ref*{fig:strategies_FP_rounds} for the FP treatment and Supplementary Figure \ref*{fig:strategies_SP_rounds} for the FP treatment.

For FP one can observe that each strategy seems to converge with time to a dominating decision-making pattern. For example, the differences we observed between sub-clusters A.0 and A.1 are determined by how they act in the first 25 rounds, where it seems participants were still exploring their options. For example, participants in sub-cluster A.1, show a mix of unconditional cooperation ($(CD)C$ = 0.292), unconditional defection ($(DC)D$ = 0.083) and conditional behaviors ($(CD)D$ = 0.125, $(DD)D$ = 0.208, $(CC)C$ = 0.167). After 25 rounds, they coordinated on mutual cooperation for the remainder of the game.  Cluster A.0 appears to have led to cooperation thanks to an initial reciprocal behavior ($(CD)D = 0.922$) that was followed by being resolving mistakes if they occur ($(DC)C = 0.922$). Something similar, but then for defection, happens in SP (see Supplementary Figure \ref*{fig:strategies_SP_rounds} ), where sub-clusters E.0 and E.1 differs essentially in how they react to contexts in the first 25 rounds, yet converge to the same behavioral pattern in the end.

The evolution of sub-cluster B.0 underlines the importance to have long enough experiments: as can be observed, these two participants started out exploring actions in the first 25 rounds, then mutually defected in the next 50 rounds, and finally found a way to cooperate in the last 25. The other two sub-clusters, B.1 and B.2 increase their probability to reciprocate defection $(DD)D$, but participants in sub-cluster B.1 showed from rounds 26-75 a willingness to establish cooperation ($(CD)C  = 0.24$ and $0.12$ respectively) and ended up acting conditionally in the last quarter of the game.

Additionally, although cluster C in FP appeared to have less well-defined behaviors associated with it, one can see that C.0 and C.2 are in fact also very specific in their HMM description.  Only for C.1 do we see many different contexts and responses in the emission probabilities of the HMM, yet this abundance of responses remains consistent over all rounds. A very similar situation happens in SP (see Supplementary Figure \ref*{fig:strategies_SP_rounds}), where even though in some clusters the strategies became more concise by having less diversity of contexts (see for instance cluster D.0). Yet, the majority of behaviors appear to remain rather stable over time, which is especially clear for cluster E.  Overall, the HMM models in the figure reveal explicitly how having new partners at each round induces noise in the decision-process and hinders participants to converge to a clear strategy, relevant for the experiences they have.

\subsection*{FP Participants organize according to their behavioral clusters}

In FP we can clearly observe that the participants either ended up in full cooperation or defection. A form of self-organization or behavioral self-selection appears to be happening (see also earlier), which was unable to establish itself in the SP treatment. To confirm this hypothesis, the inter- and intra-cluster interactions between participants are analyzed, as visualized in Supplementary Figure \ref*{fig:nets}. In panel A of that figure, the results are shown for the FP treatment: The members of each behavior cluster are always interacting in the same contextual cluster. Moreover, some behavioral clusters are completely isolated (e.g. see clusters A.0 and A.1) or interact strongly with a specific other sub-cluster in the same contextual cluster (e.g. C.0 and C.1). 

This analysis for the SP treatment shows a completely different picture: the members of one behavioral cluster interact often with members of another behavioral cluster. Even more, members of cluster E and F are more likely to interact with members of another contextual cluster than those inside their own. One can conclude here that stranger matching disrupts the self-selection process observed for FP, generating strategies that try to deal in a different manner with the situations they encounter. 

The behavioral self-selection in FP is clearly a consequence of the fixed relationships in that treatment. In that case, previous actions play a more significant role than in the case of the SP treatment. This hypothesis is confirmed by the information obtained in the short questionnaire at the end of the experiment: Each participant was asked if her decision was influenced by what the other player did in the previous round (see Supplementary Information for more details). In the SP treatment, 50 players responded that they were influenced ($52.08\%$) by the other's actions. Yet, in the FP treatment, 74 participants responded "yes" to the same question ($80.43\%$). This means that FP participant actions were shaped strongly by actions in the last round, while in SP this shaping did not take place as almost half of the participants did not care about what their opponents did, leaving them with a different approach to decide between cooperating or defect.

Yet, notwithstanding this complexity, Supplementary Figure \ref*{fig:nets} also reveals that there is some structure in the network of possible interactions between all behavioral sub-clusters found in the SP treatment. Within the same contextual cluster, E.1 - E.2 and D.0 - D.1 represent strong interaction pairs while between contextual clusters the combinations E.0 - F.0, E.1 - F.2, D.0 - E.3 and D.3 - F.1 are accentuated. These associations are stronger than the within behavior cluster associations (i.e. the loops) and many of the other possible associations. This result indicates that the clustering is finding specific, non-random, behaviors that were used by the different participants, reminiscent of some kind of strategy.

Repeated interactions with the same partner thus lead to behavioral self-selection of participants into behavioral clusters .  Similar results were previously observed in a Surplus Allocation game \cite{di_guida_repeated_2021}, wherein participants could themselves decide to continue with the same group, leading to a behavioral self-selection of groups of short and long duration, where the latter corresponded to cooperative ones.

\subsection*{Conditional strategies like TFT and WSLS are observed but not consistently used}

How do the strategies inferred from the treatment data relate to those proposed in the literature for the origins of cooperative behavior? To answer this question, the HMM results in Supplementary Figures \ref*{fig:strategies_FP_rounds} need to be considered as the theoretical strategies have been analyzed mostly in fixed partners interactions. Focusing on cluster A in FP first, one can derive that both sub-clusters in A may be associated with TFT-like or a WSLS-like behavior. Considering A.0, one can observe from round 1-25 that this cluster is associated with a reciprocal strategy: It starts out by defecting when the co-player defects  (i.e. $(CD)D$) but still cooperating when she does ($(DC)C$), leading quickly to mutual cooperation ($(CC)C$) for the remainder of the FP treatment. Given the strong presence of $(CD)D$, this behavior is almost like a form of TFT. 

Behaviors in A.1 also end up in mutual cooperation but achieve this in a different, yet less clear, manner: one could say the participants in A.1 also use a form of reciprocation when the co-player defects but cooperation appears to have been promoted while being generous ($(CD)C$ in combination with $(CD)D$) and signaling to go back to cooperation when both defect ($(DD)C$). This behavior is very much associated with the idea of WSLS (Win-Stay, Lose-Shift) or Pavlov as explained by Kraines and Kraines \cite{kraines_learning_1993}: When winning ($(DC)$ and $(CC)$ contexts) continue with the same action, when loosing ($(CD)$ and $(DD)$) switch. This makes sense in the way that this strategy was designed to promote cooperation in defective environments, and to mimic the stochasticity of biological and social interactions \cite{kraines_learning_1993}. Although non-WSLS responses are still present, it appears that together they were sufficient to induce cooperation.

This positive outcome did not occur for members of the behavioral cluster C.2, which contains a subset of participants that have similar cooperative ambitions as those in A.1 (visualized in Figure \ref{fig:context_cluster_chart}B through the proximity of a subset of C.2 participants with the A cluster), nor for members of C.1 and C0.  While the behavior of  C.2 in the first 25 rounds is very noisy, it too turns into something reminiscent of WSLS in the following window (rounds 26-50). Although C.2 participants played mostly among themselves (see Figure \ref{fig:nets}A), cooperation was not achieved given that mutual defection did not lead to cooperation ($(DD)C$).  Similarly, for C.1, the frequency of $(DD)D$ is much higher than $(DD)C$, the exploitative action $(DC)D$ is dominating, which may explain again why cooperation is not achieved (note that C.0 players that interacted with C.1 players did not follow WSLS, stimulating the exploitation by C.1 members). In SP (see Supplementary Figure \ref*{fig:strategies_SP_rounds}),  approximations of WSLS can also be observed, see for instance D.3, E.1. and E.2. Yet in all those cases $(CC)C$ is missing, again not leading to cooperation, with the stranger interaction possibly being the reason for the failure of observing that part of WSLS.

So in the case of fixed partnership interaction, we see successful cases of TFT- and WSLS-like behavior but more often the "implementations" of these strategies did not lead to cooperation, most likely because they were not rigorously applied, as opposed to the case when working with theoretical models. We also see cases of non-conditional behavior, for example in sub-cluster E.0 and E.1 in rounds 26 to 100 where the participants in these clusters defect unconditionally $(DC)D$ and $(DD)D$, which resembles the theoretical strategy AllD. The contrary also holds, where for example, sub-cluster F.0 unconditionally cooperates $(CD)C$ and $(CC)C$ during the whole game.

\section*{Discussion}

As seen in the Introduction, many approaches can be taken to study the strategies in the IPD. In this work, we clustered participants based on the context they experience. This is an important difference from other works since it allows us to see the nuances between people reacting to the same situation. We choose this approach because the act of either cooperating or defecting can mean different things depending on the context of the action. It is not the same to defect to exploit your opponent than to defect to reciprocate your opponents' actions. Moreover, it is not the same to cooperate with a fixed opponent as to cooperate with a stranger, that has been interacting with strangers as well. We used unsupervised methods which make no prior assumptions about what strategies players could use, for this, we tested several methods with the same goal to group participants with similar experiences, and we confirmed there is a significant overlap between different approaches.  

A second level clustering (behavioral clustering), on how often they cooperated is necessary to analyze the individual differences given their opponents' actions, and how they signal their intentions in each situation. In fact, in the sub-clusters we found, we could identify players that faced the same context but behaved differently. There are plenty of examples, one of them is how participants in cluster E in SP faced defection, but when presented with mutual cooperation ($CC$), players in E.2 had 20 opportunities to respond with cooperation, which only one of them did so, once in the whole game, while the participants in the same cluster and context cooperated more, while having fewer opportunities to do so. This means that to understand the strategies people use in the IPD, it is necessary to see what was happening around them in the past and observe how they play in such situations. 

Another factor that has an effect on people's strategies in the IPD is their opponent structure. Since we had two treatments, one with fixed co-players and one with shuffled ones, we could see how they act and what strategies they use given each context. Not only players in FP achieved more cooperation, but their strategies looked similar to their co-players. In other words, they were better at self-organizing and the intra-cluster behavior was very similar, as opposed to players of the SP treatment, who had to interact with members of other sub-clusters, who at the same time, had different strategies in mind. 

One last hypothesis that we were able to prove here was the learning effect on people's strategies in the IPD. By dividing the analysis into windows of 25 rounds, we could see that the initial quarter was used by the players to explore and try different options, we could even see this effect in a chaotic environment such as SP, where establishing an action plan can be difficult due to its opponent structure.

Using Hidden Markov Models proved to be a handful tool to visualize the strategies and approximate how these underlying factors (in this case, represented as hidden states) shape participants' behavior. Taking a probabilistic approach, rather than a rigid, deterministic one accounts for the stochastic nature of our interactions. 

Our findings have implications on how strategies in the IPD are studied: first, performing long experiments allowed us to observe that the participants may not have one but many strategies throughout the game before reaching an equilibrium. This means that when we talk about "how do we really act" we need to account for an exploration stage and a stabilization point. Second, we found that the opponent structure has an effect on people's behavior, and therefore, their strategies. Whether they are more adaptive to their opponents' actions or persistent based on their initial expectations and preferences, this depends on how the game is designed, different strategies might arise. For this reason, we believe that a logical step towards developing these results further is to analyze these strategies in other scenarios, such as N-player IPD, as opposed to the pairwise setting we presented in this paper. Previous work has stated that cooperation could be hindered \cite{grujic_three_2012}, but it is uncertain whether they will still be able to reach an equilibrium in their strategies as fast as the pairwise game, and from the strategic point of view, their rationale of what constitutes exploitation or reciprocation changes in this case.

Although we could observe some identifiable strategies thanks to the theoretical work behind the IPD, it is still challenging to determine and predict human behavior in these games, and this was represented by the big diversity of options on each HMM. And while taking a probabilistic approach to model decision-making might help us to understand our heuristics, this will continue to be an open issue in research on human cooperation, since all models have their limitations and there are many other scenarios to take into account. Another factor that is well known is the noise that not only our actions produce, but also our perception of what others are doing and their expectations, which they may also have about us. This is a caveat of human interaction to overcome and we are still trying to understand through these experiments.

\bibliographystyle{ieeetr}
\bibliography{references}  





\section*{Ethics}
Ethical approval by reference number ECHW2015\_3 was obtained from the Ethical Commission for Human Sciences at the Vrije Universiteit Brussel to perform this experiment. All the participants in the study had to give their informed consent prior to the participation.

\section*{Acknowledgements}

E.M. and T.L benefit from the support by the Flemish Government through the AI Research Program and by TAILOR, a project funded by the EU Horizon 2020 research and innovation program under GA No 952215.  E.F.D is supported by an F.N.R.S Chargé de Recherche position, grant number 40005955. J.G. was supported by the FWO - Research Foundation Flanders. T.L. is furthermore supported by the F.N.R.S. project with grant number 31257234,  the F.W.O. project with grant nr. G.0391.13N, the FuturICT 2.0
(\url{www.futurict2.eu}) project funded by the FLAG-ERA JCT 2016 and the Service Public de Wallonie Recherche under grant n\textdegree\ 2010235–ARIAC by DigitalWallonia4.ai.   
\section*{Author contributions statement}

J.G. designed, performed the experiments, and collected the data. E.M., J.G, T.L., and E.F analyzed the results. E.M. and J.L. performed the clustering and developed the models. E.M., E.F., and T.L. wrote the manuscript. All authors read and approved the manuscript.

\section*{Additional information}

\subsection*{Competing interests}

The authors declare no conflict of interest. The funding agency had no role in the design of the study; in the collection, analyses, or interpretation of data; in the writing of the manuscript, or in the decision to publish the results.

\subsection*{Data availability}

The experimental data, code used to fit the models, plot the figures, and other supplementary information, is available at \href{https://datadryad.org/stash/share/YpoJwOsFpYJGKV0qS8MAyqVpJ7C9Vp13eQ16Bca-jGg}{doi:10.5061/dryad.37pvmcvmk}.

\newpage

\section*{Supplementary Information}

\newpage

\begin{table}[!ht]
\centering
\begin{tabular}{|c|c|c|c|c|}  
\hline
\textbf{Treatment} &\textbf{Context previous round} &\textbf{Frequency} & \textbf{Cooperation} & \textbf{(\%)}\\
\hline
Fixed Partners & CC & 4,122 & 3,910 & 77.38\% \\ 
 & CD & 980 & 562 & 57.34\%\\ 
 & DC & 980 & 320 & 32.27\%\\ 
 & DD & 3,026 & 316 & 10.44\%\\ 

\hline
Shuffled Partners & CC & 806 & 647 & 80.27\% \\ 
 & CD & 1,910 & 1,127 & 59.0\%\\ 
 & DC & 1,910 & 290 & 15.18\%\\ 
 & DD & 4,878 & 626 & 12.83\%\\ 

\hline
\end{tabular}
\caption{Frequency table for each context in the previous round and the percentage of cooperation given the context, per treatment. }
\label{table:context}
\end{table}

\clearpage

\clearpage

\clearpage

\section*{Modularity Network} \label{methods:mod_net}

As mentioned in the Methods section of the main paper, we performed different clustering methods to test the separation between participants and their behavior. In this case, we also used the Modularity Network (MN) clustering, especifically the Louvain method \cite{blondel_fast_2008}. In our case, a network was built, where each node represented a subject on each treatment (FP and SP) and they were connected by a weighted edge. We used eight variables representing the context and the cooperation under each context used in the other clustering methods (K-Means and Hierarchical Clustering), represented by a vector $S$:

\begin{equation}
    S = ((CC), (CD), (DC), (DD), (CC)C, (CD)C, (DC)C, (DD)C)
\end{equation}

This way, the distance between participant $i$ and participant $j$ is: $d = ||S_i - S_j||$. The weight between the nodes is given by the inverse of this distance plus one, to avoid divisions by zero: 

\begin{equation}
\centering
w = \frac{1}{d + 1}    
\end{equation}

\begin{figure}[!ht]
\centering
 \includegraphics[width=0.90\linewidth]{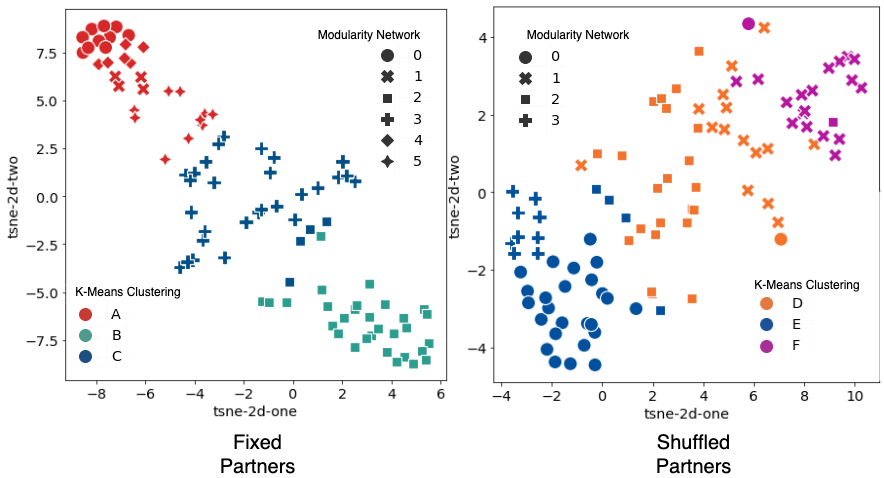}
\caption{tSNE Plot Modularity Network (MN) and K-Means clustering. As seen in the previous figure, the colors are the clusters found with K-Means and the shapes the clusters found with MN clustering. The overlap is still visible, even though they are two different approaches to unsupervised learning.}
\label{fig:modnet}
\end{figure}

In MN clustering, one aims to maximize the modularity M, in this case, our clustering resulted in $M=0.40$ for all treatments. To test the robustness of the clustering a randomized network was built over 100 iterations and the same process of getting the modularity was done, this random network resulted in $M=0.264\pm0.006$. Figure \ref{fig:modnet} shows the overlap between the K-Means and the MN clustering.

\clearpage

\begin{figure}
\centering
\includegraphics[width=0.8\linewidth]{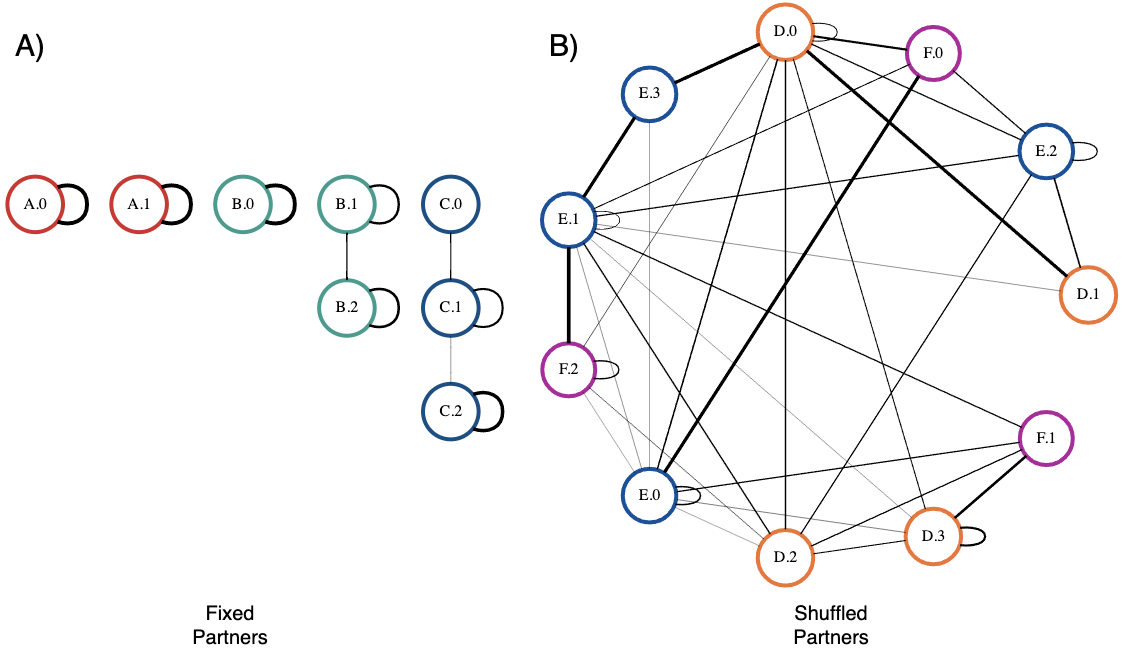}
\caption{Network showing the interaction between sub-clusters in both treatments. \textbf{A)} FP treatment and \textbf{B)} SP treatment. In SP, it can be seen how subclusters interact with members of other clusters, in contrast, a self-organisation can be seen in FP since members of the same sub-cluster interact with each other, or in the case of sub-clusters B.1, B.2, C.0, C.1 and C.2 interact with members of the same cluster (B and C respectively).}
\label{fig:nets}
\end{figure}

\clearpage

\begin{figure}[!ht]
\centering
\includegraphics[width=\linewidth]{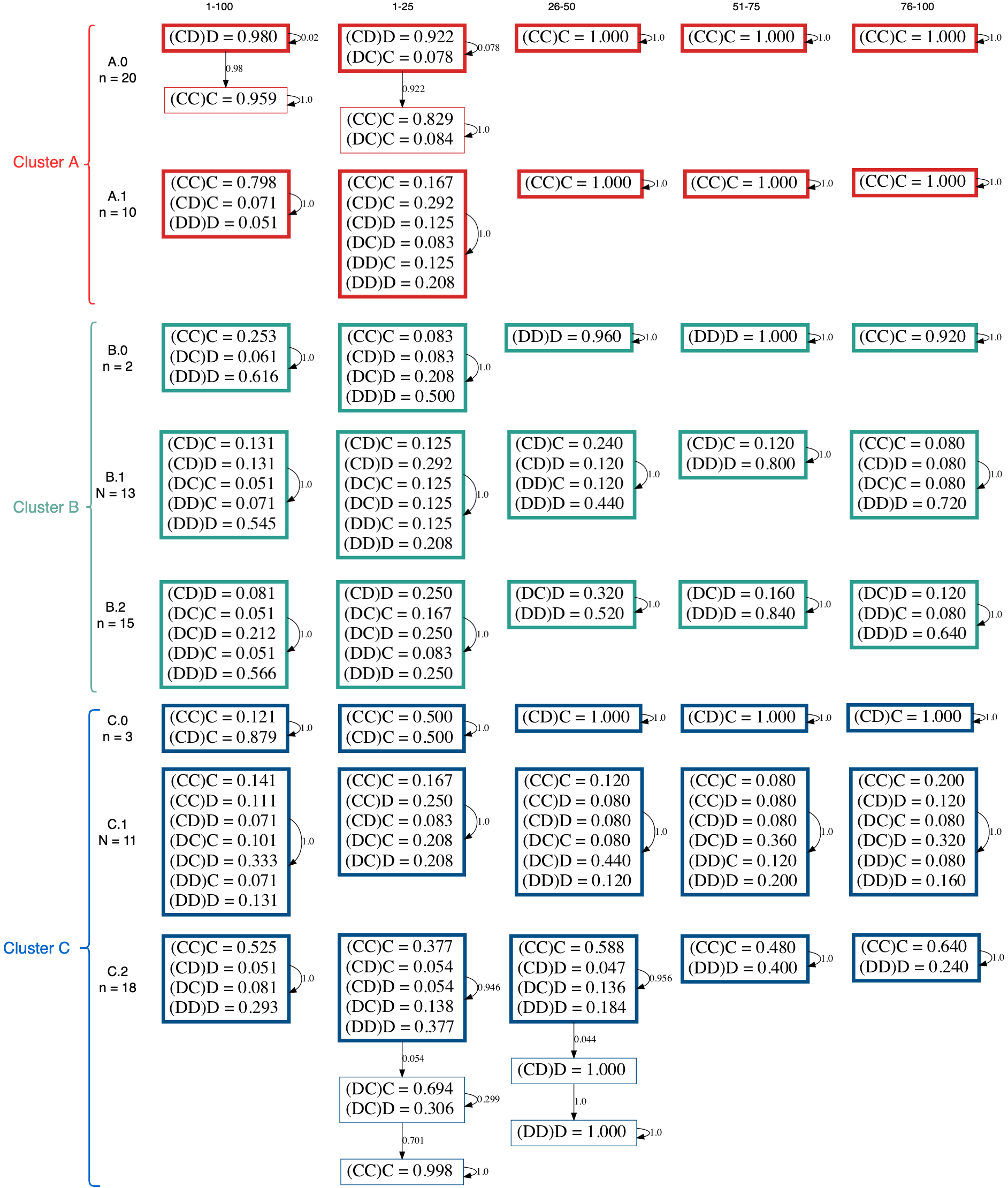}
\caption{Hidden Markov Models for the Fixed Partners treatment. Here, the eight sub-clusters found in FP have a HMM that describes each sub-cluster's strategy. Bold rectangles represent the initial state, while the others represent subsequent hidden states. At the top, round windows are specified, starting from 1-100, then in windows of 25.}
\label{fig:strategies_FP_rounds}
\end{figure}

\clearpage

\begin{figure}[!ht]
\centering
\includegraphics[width=0.8\linewidth]{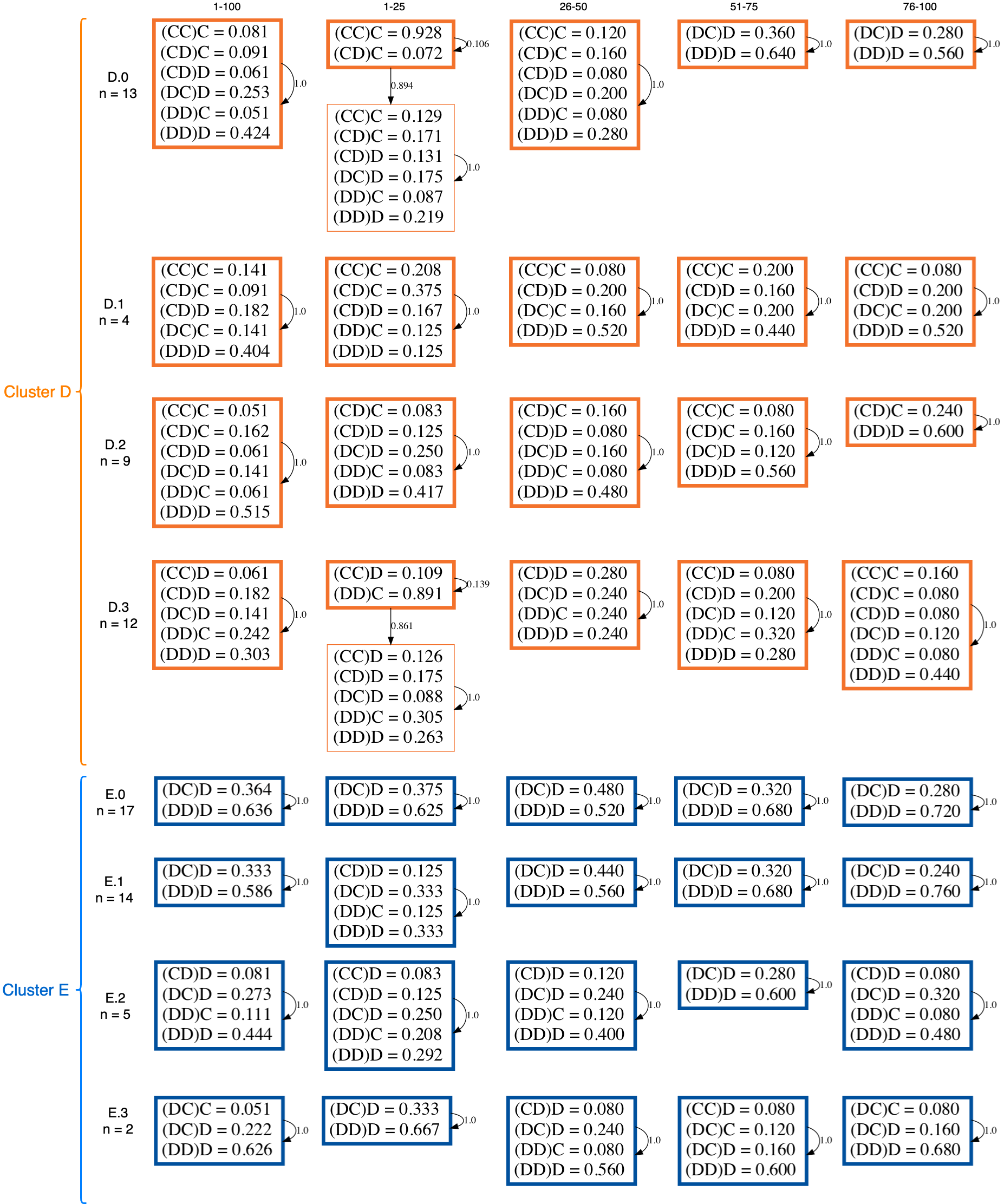}
\end{figure}

\end{document}